\documentclass[twoside,12pt]{article}
\usepackage{epsfig,latexsym}

\def\Journal#1#2#3#4{{#1} {#2} (#4) #3 }

\def\PHYS{{\em Physica}}

\def\NPB{{\em Nucl. Phys.} B}

\def\PLB{{\em Phys. Lett.} B}

\def\PRL{\em Phys. Rev. Lett.}
\def\PREV{\em Phys. Rev.}
\def\PREP{\em Phys. Rep.}
\def\PRA{{\em Phys. Rev.} A}
\def\PRD{{\em Phys. Rev.} D}

\def\ANNP{\em Ann. Phys. (N.Y.)}
\def\RMP{{\em Rev. Mod. Phys.}}

\newcommand{\be}{\begin{equation}}
\newcommand{\ee}{\end{equation}}
\newcommand{\bea}{\begin{eqnarray}}
\newcommand{\eea}{\end{eqnarray}}

\newcommand{\ub}[1]{\underline{#1}}

\newcommand{\Pminus}{{\cal P}^-}

\def\senk#1{\vec{#1}_\perp}

\begin{document}

\title{ \vspace{1cm} Casimir Force on a Light Front}

\author{S.S.\ Chabysheva,$^1$ J.R.\ Hiller$^{1,2}$\\
\\
$^1$Department of Physics, University of Idaho, Moscow, ID 83843 USA \\
$^2$Department of Physics and Astronomy, \\
University of Minnesota-Duluth, Duluth, MN 55812 USA}

\maketitle

\begin{abstract} 
Depending on the point of view, the Casimir force arises from variation in
the energy of the quantum vacuum as boundary conditions are altered or as
an interaction between atoms in the materials that form these
boundary conditions.  Standard analyses of such configurations are
usually done in terms of ordinary, equal-time (Minkowski) coordinates. 
However, physics is independent of the coordinate choice, and an analysis
based on light-front coordinates, where $x^+\equiv t+z/c$ plays
the role of time, is equally valid.  After a brief historical
introduction, we illustrate and compare equal-time and light-front 
calculations of the Casimir force.
\end{abstract}

\section{Introduction} \label{sec:Intro}

The archetype of the Casimir force~\cite{Casimir,Plunien,MilonniText,%
Mostepanenko,Dittrich,Milton:2001yy,Milton:2004ya,%
LamoreauxReview,Simpson} is the interaction between
two closely spaced parallel conducting plates.  A loose 
argument for the force is that the plates restrict
the number of permitted electromagnetic modes between
the plates while the space outside enjoys the full
range; the difference in zero-point energy densities causes an 
inward pressure.  Though intuitive, the argument fails 
in other situations and is actually misleading, because both 
densities are infinite and require regularization.  For
example, the same argument would imply that a conducting
spherical shell would experience a collapsing, inward force,
but this is not the case.\footnote{There was an attempt
to balance this supposed inward force by the outward
repulsion of the electron's charge density and arrive
at a mechanical model of the electron~\protect\cite{electron,Boyer}.}

Standard analyses of such configurations are
usually done in terms of ordinary, equal-time (Minkowski) coordinates. 
However, physics is independent of the coordinate choice, and an analysis
based on light-front coordinates~\cite{Dirac,LFreview1,LFreview2,LFreview3,%
LFreview4,WhitePaper,LFreview5}, where $x^+\equiv t+z/c$ plays the role of time, 
is equally valid.  This review provides a survey of applications of
light-front coordinates to the Casimir force~\cite{Lenz,Almeida,LFCasimir,Rodrigues}
and compares the results with equal-time calculations for a massless scalar field
with boundaries fixed by parallel plates.  Extensions to other fields, such as
photons, is straightforward.  The primary complication that arises for
light-front calculations is for plates perpendicular to the 
longitudinal direction, for which the exact nature of the
boundary conditions relative to the light-front space and time
coordinates is critical.

The electromagnetic Casimir effect can also be described
in terms of the van der Waals interaction between the
atoms in the boundary materials.  However, the effect can
occur in principle for any quantum field, while the
van der Waals force is limited to electromagnetic 
interactions.  The only question is how to select
appropriate boundary conditions for the particular field.
Implicit in the boundary conditions will be a force,
analogous to the van der Waals force, between the elements
of the materials forming the boundaries.

The reality of the vacuum energy is supported by a
demonstration that the inertial and gravitational
masses of Casimir energy are proportional~\cite{Fulling},
consistent with the general relativistic equivalence principle.
However, the interpretation that the Casimir effect provides proof that
vacuum energies are real has been questioned~\cite{Jaffe}.
Because the Casimir force can in principle be computed
from the van der Waals force between atoms in the material forming
the boundaries, the vacuum energy does not need to be invoked to
explain the Casimir force.  However, from the point of view
of assembling a configuration that gives rise to a measurable
Casimir force, one can compute the work done during an adiabatic assembly
and translate that into a potential energy, the gradient of
which determines the force acting between the assembled
parts.  One would normally consider this potential energy to
be stored in a rearrangement of the electromagnetic
field, which in the case of the Casimir effect, would be
vacuum modes.  An analogy is the energy stored in a configuration
of two charges, brought together against the Coulomb force
between them; this energy is associated with the energy
of the electrostatic field minus the self energies.  The two
views are complementary.

Vacuum effects such as the Casimir force and Casimir torque will
be important for the design of nanoscale devices.  The
interatomic forces can create surface adhesion effects
that can even break such devices.  Conversely, the effects
can be incorporated to improve sensitivity of accelerometers
and torsion sensors~\cite{Munday}.

The remainder of this paper contains a brief historical overview in
Sec.~\ref{sec:history}, followed by specific calculations of
the Casimir force between parallel plates in Sec.~\ref{sec:calculations}.
The calculations compare use of light-front coordinates with
that of equal-time coordinates for the case of a massless
scalar field subject to either periodic boundary conditions
or zero boundary conditions at the plates. For the light-front
case, plates are considered both parallel and perpendicular to the 
longitudinal direction.  A summary is included as Sec.~\ref{sec:summary}.

\section{Historical background} \label{sec:history}

\subsection{\it Precursors}

In classical physics, one can obtain an attractive
$1/r^6$ potential between polarizable objects a distance $r$ apart.
This is driven by thermal fluctuations that induce
a dipole moment in one object which in turn induces a dipole
moment in the other.  The effect is linear in the temperature $T$.
This is contradicted by quantum mechanics where the
effect is constant at low $T$ and vacuum fluctuations
play the role of the thermal fluctuations.

The original proposal by van der Waals for an attractive force between
molecules was associated with a Yukawa type potential $-A\frac{e^{-Br}}{r}$
that was to explain the pressure term in the equation of state
\be
(P+\frac{a}{V^2})(V-b)=nRT.
\ee
It was London who obtained the correct form for short distances~\cite{London}
\be
V=-\frac{3\hbar\omega_0\alpha^2}{4r^6}
\ee
from fourth-order perturbation theory, with $\omega_0$ the characteristic
transition frequency for transitions to excited states
and $\alpha$ is the static polarizability.

Casimir and Polder improved on this result for larger distances
where $\omega_0 r/c$ is much greater than unity and retardation
becomes important.  The London result is then modified to~\cite{CasimirPolder}
\be
V=-\frac{23\hbar c}{4\pi r^7}\alpha_A\alpha_B
\ee
for a pair of atoms with polarizabilities of $\alpha_X$.  This result was
initially computed by two-photon exchange in perturbation theory but
is consistent with polarizations induced by vacuum fluctuations~\cite{CasimirChim}.

In 1947, Lamb and Retherford measured the $2S_{1/2}$-$2P_{1/2}$
split in atomic levels~\cite{Lamb}.  This is usually calculated
as a perturbation due to virtual photon exchange~\cite{Bethe}.
However, as early as 1948, Welton interpreted the energy shift
as a consequence of interaction with the vacuum fluctuations
of the electromagnetic field~\cite{Welton}.  The vacuum field
fluctuations induce fluctuations in the electron position,
essentially a Stark effect driven by the vacuum field,
which cause corrections to the Coulomb potential; these 
corrections result in the Lamb shift.  It was also 
suggested by Feynman that the Lamb shift could be 
considered as a change in the zero-point field energy
due to the presence of atoms~\cite{Feynman}.  The 
Lamb shift is also modified by the presence of a nearby surface,
which alters the available photon modes.

The relationship between fluctuation and
dissipation is such that if a system can dissipate
energy to its surrounding reservoir irreversibly, 
then the reservoir must induce fluctuations in
the system.  Thus radiation reaction implies
the existence of a vacuum field.
Consequently, the spectral density of the vacuum field must be
directly related to the form of the radiation
reaction.  Because the radiation reaction is
proportional to the third derivative of the position,
the spectral energy density of the vacuum is
proportional to the third power of the 
frequency~\cite{CallenWelton,Milonni}.
Dissipation due to radiation reaction will
act to exponentially damp the position which would
drive the fundamental
commutator $[x,p]$ of position and momentum to zero;
noncommutativity is preserved by the fluctuations of
the vacuum.

The concept of zero-point energy began with Planck in 1912,
with his ``second'' theory of blackbody radiation.
Heisenberg developed the concept in connection 
with a quantum harmonic oscillator.  From a more
modern, field-theoretic viewpoint, the electric and 
magnetic field operators do not commute, and,
therefore, they cannot both be determined simultaneously.
Consequently, the energy of the electromagnetic state,
which has contributions from both fields,
can never be zero.  Zero-point fluctuations in crystals
do scatter light at zero temperature, so much so that
for helium that it cannot crystallize and x-ray
diffraction is impacted in ordinary crystals.
A shift in energy, to remove the zero-point energy from
the Hamiltonian, does not eliminate vacuum effects.
In particular, the expectation values for the momentum
squared and position squared remain nonzero in the vacuum
and fluctuations in momentum and position remain.

W.~Pauli rejected the notion of zero-point energy, saying
``it does not create any gravitational field, as is known
from experience''~\cite{Rechenberg}.  This is to be 
contrasted with the more recent puzzle, that the 
contribution of zero-point energy to the cosmological
constant is too large by 120 orders of magnitude~\cite{Weinberg}.
Such contradictions aside, the Casimir effect can play
a role in the structure of space-time.  It can provide
a source for compatification of extra dimensions~\cite{Candelas}
and, more generally, it can factor into the vacuum
energy of space-times that are not flat~\cite{LachiezeRey}.
Measurements of the local vacuum energy density could then,
in principle, constrain the global topology of the 
Universe.

\subsection{\it Casimir effect}

The attractive force between two conducting plates was
first calculated by Casimir in 1948~\cite{Casimir}.
He was apparently not familiar with the work of Lamb
or Welton but was instead motivated by a remark by
Bohr about zero-point energy.  The original Casimir-effect paper
cites his own work on the retarded van der Waals interaction that 
included zero-point fluctuations.  An alternate 
derivation, based on interactions of the charges in
the plates rather than the vacuum energy, has been
given by Schwinger, DeRaad and Milton~\cite{Schwinger}.

The basic result is that, for parallel plates separated
by a distance $a$ with surface area $A$, the leading 
$a$-dependent contribution to the vacuum energy between 
the plates is
\be
E_0(a)=-\frac{\pi^2 \hbar c A}{720 a^3}.
\ee
The construction requires regularization of the sum
over zero-point modes but is independent of the
regularization.  Terms independent of $a$ are subtracted
as self-energies associated with individual plates.
The Casimir force per unit area is then
\be
P=-\frac{1}{A}\frac{dE_0}{da}=-\frac{\pi^2\hbar c}{240 a^4}.
\ee
The presence of $\hbar$ in both the energy and the force
shows this to be a quantum effect.  Typical magnitudes are
$P=10^5$ Pa for $a=10$ nm and $P=10^{-3}$ Pa for $a=1$ $\mu$m.

Calculations of the Casimir effect usually rely on separation
of variables in the wave equation to obtain the eigenspectrum,
which requires simple symmetric geometries for boundaries.
The result for parallel plates can be obtained as a limit
of a calculation for a wedge~\cite{Mostepanenko}.

Instead of impenetrable walls with boundary conditions of
zero, periodic boundary conditions can also be considered.
For a massless scalar field in one space dimension with
periodicity $a$, the energy density is $-\pi/6a^2$, to
be compared with $-\pi/24a^2$ for impenetrable walls
separated by $a$.

Casimir's result was extended to realistic materials by
Lifshitz, who derived the Casimir force between plates
with complex dielectric permittivity $\epsilon(\omega)$~\cite{Lifshitz}.
Casimir's result is recovered in the limit of $\epsilon\rightarrow\infty$.
The determination requires consideration of the interior of
the media as well as the gap between.  The interactions
are not additive, because of screening effects that weaken
the van der Waals interactions between individual atoms.  The 
Casimir energy and force are then most 
conveniently given in terms of the permittivity and permeability 
of the media. An alternative is to represent the media effects
through an impedance at the boundary.  The treatment of the 
media is justified by the Ewald-Oseen extinction theorem~\cite{BornWolf},
which limits significant field penetration to approximately one
wavelength.  The $a^{-4}$ power-law behavior remains, but
the coefficient is different.  The screening effect can be neglected 
only if the media are dilute.  Penetrable walls can be modeled with 
delta functions~\cite{Jaffe}.

Because a calculation based on the van der Waals force
shows that the Casimir force
is dependent on the magnitude of the fine-structure constant $\alpha$,
there is also criticism~\cite{Jaffe} of the standard Casimir result
for being independent of $\alpha$.  It is obtained from more sophisticated
calculations only in the limit of
infinite $\alpha$.  However, the standard
result is derived for perfectly conducting surfaces, which is
equivalent to taking an infinite value for $\alpha$ from the
beginning of the calculation; the boundary
conditions correspond to the zero skin depth associated with
infinite conductivity, and $\alpha$ is proportional to the 
conductivity~\cite{Jackson}.

Other geometries and materials can result in a cutoff dependence
which reflects the physical properties of the materials at
high frequencies.  For conducting plates, the surfaces become
effectively transparent at high frequencies; however, the dominant contribution
to the Casimir force comes from frequencies well below 
this cutoff.

For a spherical conducting shell of radius $a$, Boyer obtained
a Casimir energy of~\cite{Boyer} (See also \cite{Schwinger,Bowers}.)
\be 
E_0(a)=+\frac{0.09\hbar c}{2a}.
\ee
This corresponds to a repulsive force that tries to expand
the sphere, but a negative result and an attractive force
can be obtained if one
considers realistic material for the shell~\cite{Barton}. 
Kenneth and Klick~\cite{Kenneth} also show that the Casimir force
between two dielectrics must be attractive, which applies
here because the shell can be cut into two hemispheres
which must then attract.

The evaluation for a spherical shell requires some 
care~\cite{Mostepanenko} because of
surface divergences that require additional counterterms
in the case of a massless scalar field.  However, for the
electromagnetic field, the inside and outside surface
divergences cancel, in the principal-value sense, and
one gets a finite result after simply subtracting the
free-space density.

When the boundaries are asymmetric or made from
materials that are anisotropic, such as a
dielectric birefringent crystal, there is a Casimir 
torque as well as a force.  The optical axis with
the highest refractive index tries to align, resulting
in a torque.  A torque measurement has been done recently by Somers and
co-workers~\cite{Somers}.

\subsection{\it Related phenomena}

Placing an atom in a bounded domain causes energy shifts
to depend on the nature of the normal modes for the given
boundary conditions and on the position of the atom.  This
can result in a force between a boundary and the atom.
This can include retardation effects that change the
dependence on separation $r$ from $r^{-3}$ to $r^{-4}$
as the atom interacts with its image in the conducting
plate. Casimir and Polder explained this result in terms of 
zero-point energy~\cite{CasimirPolder}.  It can also be obtained
by a modification of the Lamb-shift analysis where the
virtual photon is subject to the boundary conditions
associated with the plate~\cite{Simpson}.  The spectroscopy
of atoms between two spherical mirrors has been considered
by Heizen and Feld~\cite{Heizen}.

The rate of spontaneous emission from an excited atom is 
modified by the presence of a reflecting surface.  This
has been verified experimentally by Drexhage~\cite{Drexhage}.
In particular, placement of an atom between two reflecting
mirrors can suppress spontaneous emission when the mirror
spacing is less that one-half of the wavelength of the 
light emitted in the transition, because the necessary
mode is not available~\cite{Barton2,Knight,Philpott}.
This has also been observed experimentally~\cite{Hulet}.
In general, the discrete spectrum of a confined space
may not include a mode that resonates with a particular
transition frequency.

Spontaneous decay is driven by interaction with the
vacuum photon energy density and by radiation reaction
due to fields generated by fluctuations in the 
atomic dipole moment. The interaction with the
vacuum photons is analogous to stimulated emission
where the atom interacts with thermal photons;
however, this interaction accounts for only half
of the spontaneous emission rate, with the other half
due to radiation reaction.  For the ground state,
the two effects cancel, and there is no
spontaneous absorption.

Magnetic moments are affected by boundary conditions.
For example, measurements of the anomalous magnetic
moment $g-2$ for the electron that use a Penning trap
must take into account the consequences of cavity
QED, which is to introduce an interaction with
the vacuum fields of the cavity.  From the point of view
of perturbation theory, the photons in loops must occupy
states allowed by the cavity, which is a restriction
from free-space modes.  Welton attempted a derivation
of $g-2$ based on interactions with the vacuum fields
but obtained the wrong sign~\cite{Welton}.  This was
corrected to include radiation reaction effects that
renormalize the mass~\cite{Grotch}; however, the
result remains cut-off dependent.

The bag model for hadrons~\cite{BagModel} includes contributions
to the energy from the Casimir effect for interior quarks
and gluons, trapped in a spherical ``bag.''  The
effect is about 10\% of the mass for a nucleon, despite that
fact the the free-space zero-point energy for fermions is
negative.  The computation of the Casimir energy~\cite{Milton}
typically ignores interactions between the
quarks and gluons.  For the quarks, the boundary conditions cannot
be simply the setting the Dirac field to zero; instead
a condition is imposed that prevents flux through the
boundary.  This yields an attractive force for fermions
between plates and a repulsive force for a spherical
shell~\cite{Chodos,Johnson,MiltonAnnPhys}.

\subsection{\it Regularization}

In a quantum field theoretic analysis of vacuum effects,
the choice of operator ordering can determine whether an
effect is due to vacuum fluctuations or radiation reaction
or a combination of the two.  For spontaneous emission, 
there is no choice of ordering the will allow for the
effect to be purely due to vacuum fluctuations; for
standard normal ordering, the rate is determined by
radiation reaction, and for antinormal ordering by
a combination with vacuum fluctuations~\cite{Milonni,Smith}.
For the Lamb shift, the effect is due to vacuum
fluctuations for symmetric ordering, radiation reaction
for normal ordering, and a mixture for antinormal 
ordering~\cite{Smith,Ackerhalt}.  The ordering of
creation and annihilation operators does, of course,
control vacuum effects to some extent; normal ordering
explicitly subtracts them from the Hamiltonian.  From
this point of view, the Casimir force can be attributed
to source fields when normal ordering  is used and to
vacuum fields when symmetric ordering is used.

Computation of the Casimir force from vacuum energy 
requires subtraction between two infinite quantities, 
the vacuum energy in free space and the vacuum energy 
in the bounded space.  The leading divergence of the 
free vacuum energy is of order $p^d$, where $p$ is the 
cutoff momentum and $d$ is the dimension of space-time.
However, the free vacuum energy is computed as an integral, and
the bounded-space vacuum energy as a sum.  The simplest approach, as done
originally, is to introduce cutoffs, to be removed
after the subtraction.  However, cutoff independent
methods have been developed, based on Greens functions
and on the Abel-Plana formula~\cite{AbelPlana,Mostepanenko}
\be
\sum_{n=0}^\infty F(n)-\int_0^\infty F(t)dt
  =\frac{F(0)}{2}+i\int_0^\infty dt \frac{F(it)-F(-it)}{e^{2\pi t}-1}.
\ee

There are actually two ways to subtract the free-space
energy.  One is to work with the boundary conditions
already in place, and the other is to bring the boundaries
in from infinite separation.  The second approach is
useful when a boundary itself is associated with an infinite
energy relative to free space; this energy must be
subtracted along with the free-space energy.  This happens
because there can exist surface divergences in the
vacuum energy density that are nonintegrable, making the
energy associated with the vicinity of the surface infinite,
even for a compact domain.  This can be controlled by
subtraction, as already mentioned, or by invoking more
realistic boundary conditions that do not create the
density divergence in the first place.

\subsection{\it Measurements}

Early tests of the Casimir effect were difficult to do and
inconclusive.  The first qualitative observation was by
Sparnaay in 1958~\cite{Sparnaay}, where the results are
described as being ``not inconsistent'' with the prediction
by Casimir.  The first quantitative measurement was
done by Lamoreaux~\cite{Lamoreaux}, using a sphere and
a plate rather than two plates, to avoid the difficulty
associated with maintaining plate alignment; it confirmed
the Casimir force (as modified for a spherical surface)
for separations between 0.6 and 1 $\mu$m.  For shorter
separations, the Casimir force becomes just the unretarded
van der Waals-London force; for larger, thermal effects
become important.  Subsequently, there were many experiments
testing various aspects, including geometry, finite
temperature, and material properties.  In addition,
the direction of the force has been observed to reverse
from attractive to repulsive when a
dielectric is placed between objects~\cite{MundayCapasso}.
For a recent citation of experiments and discussion of methods,
see \cite{Simpson}.  

For a list of early experiments intended
to confirm the existence of zero-point energies, see \cite{Clausius}.
In particular, Mulliken showed in 1924, before Heisenberg's
derivation of zero-point energy, that B$_{10}$O$_{16}$ and 
B$_{11}$O$_{16}$ molecules have a nonzero minimum vibration 
energy.  A much more recent experiment shows evidence of a lower bound on
the motion of laser-cooled trapped ions~\cite{Diedrich}.

Some common features of the Casimir-force experiments include measurement
of the force with a mechanical transducer, via its change in position
or in resonant frequency; calibration with electrostatic forces;
vibration isolation; minimization and strict accounting of
external forces between the objects and with surroundings,
including electrostatic, magnetic and gravitational forces;
measurement and control of separations with screws
and piezoelectrics.  Surfaces are typically gold
plated; however, measurements with other metals, such
as copper, and semiconductors, such as germanium and
silicon, have been done.  Imperfections in the materials
can have a significant effect.  

Most experiments are done
at room temperature.  At low temperatures, noise
caused by refrigeration equipment becomes a serious
problem; however, low temperature does reduce 
electron-phonon scattering in the materials, and
results indicate that the Casimier effect is 
independent of these interactions.  High temperatures
are impractical due to thermal expansion of the
apparatus.  

Some remaining experimental issues
include having a model for the frequency dependence
of the permittivity, in order to compare with theory,
and residual electrostatic interactions associated
with surfaces not being exact equipotentials.  Use
of atomic force microscopy methods, with spherical
or cylindrical probes, may be useful.

All of these measurements focus on a Casimir force 
derived from electromagnetism.  There has not been
a measurement of a Casimir force due to any other
type of field.  However, measurement of the electromagnetic
Casimir force can be used to constrain models
of new long-range forces due to exchange of
light particles~\cite{FeinbergSucher,Sokolov}
as well as search for deviations
from Newtonian gravity.

A related experiment~\cite{Lundeen} tests the polarization of 
an atomic core by an electron in a high Rydberg state.
Bernab\'{e}u and Tarrach~\cite{Bernabeu} had applied
dispersion theory to obtain a retardation correction 
\be
V=\frac{11\hbar e^2\alpha}{4 \pi m c r^5}
\ee
to the potential for an electron a distance $r$ from a polarizable 
object with polarizability $\alpha$.
Kelsey and Spruch~\cite{Kelsey} obtained the same
correction using perturbation theory.  They also derived
this form by invoking the zero-point energy of the electromagnetic
field~\cite{Spruch}.
These results were confirmed in a
dispersion theoretic analysis by Feinberg and 
Sucher~\cite{Feinberg}.

\subsection{\it Thermal effects}

The spectral density of photons at a temperature $T$ is
\be
\rho(\omega)=\frac{\hbar\omega^3}{\pi^2 c^3}\left[n(\omega)+\frac12\right],
\ee
where $n(\omega)=1/(e^{\hbar\omega/kT}-1)$ is the
thermal photon number in Bose-Einstein statistics.
The $\frac12$ term is from virtual photons, and the limit of
$T\rightarrow0$ yields the spectral energy density of
the vacuum
\be
\rho_0(\omega)=\frac{\hbar\omega^3}{2\pi^2 c^3}.
\ee
This is consistent with the requirement that the vacuum
spectral density be Lorentz invariant~\cite{Boyer2};
uniform motion relative to the vacuum cannot be detected.
The requirement of an $\omega^3$ dependence follows from
the force on a neutral system moving through a 
thermal radiation field with spectral density $\rho$,
which is given by~\cite{Hopf} $\rho-\frac{\omega}{3}\frac{d\rho}{d\omega}$
and is zero for $\rho\propto\omega^3$.  

For an accelerated observer, there is an effect; the observer sees 
the equivalent of a thermal bath with an effective temperature~\cite{Unruh}
$T=\hbar a/2\pi kc$, with $a$ the acceleration.  If the external
force an a charged object is constant, the acceleration will fluctuate
as the charge interacts with the fluctuations in this apparent
thermal bath of vacuum fields.  The charge then radiates, which
damps the motion~\cite{Sciama}.

Thermal effects for the Casimir force become important when the 
separation is larger than the thermal wavelength $\hbar c/kT$.  The
force between plates is then dominated by thermal photons
with a pressure of~\cite{Schwinger}
\be
P=-\frac{2.4 kT}{4 \pi a^3}.
\ee
This has a $a^{-3}$ dependence instead of $a^{-4}$ and
is independent of $\hbar$, making it no longer a quantum
effect.  At smaller separations, experiments need to
take thermal effects into account in order to
accurately evaluate the Casimir force.  For finite temperature,
the zero-point energy $E_0=\frac12\sum_n \hbar\omega_n$
is augmented as the free energy~\cite{Plunien}
\be
{\cal F}=\sum_n\left[\frac12\hbar\omega_n+kT\ln(1-e^{-\hbar\omega_n/kT})\right].
\ee
For temperatures much below $\hbar c/4\pi ka$, the pressure
between parallel plates is~\cite{Levin,Schwinger}
\be
P=-\frac{\pi^2\hbar c}{240a^4}\left[1+\frac13\left(\frac{T}{T_{\rm eff}}\right)\right],
\ee
where $T_{\rm eff}\equiv \hbar c/2ak\sim10^3$K $\mu{\rm m}/a$.  Thus,
for separations on the order of a micron, the effective temperature
is quite high and temperature effects are low.  The analysis assumes
that the apparatus is at thermal equilibrium; for surfaces at different
temperatures, thermal photons are emitted and absorbed at different
rates.  For realistic
materials, there is also temperature dependence through the
permittivity and the permeability.

\section{Calculations} \label{sec:calculations}

\subsection{\it Equal-time quantization} \label{sec:ETquant}

For simplicity, we present the calculation for a free massless scalar field $\phi$, subject
to boundary conditions consistent with a chosen positioning of the plates.
The Lagrangian is\footnote{In this section, we use units for which $\hbar=1$ and $c=1$.}
\be
{\cal L}=\frac12(\partial_\mu\phi)^2,
\ee
and the Hamiltonian density is
\be
{\cal H}\equiv \partial_0\phi\partial_0\phi-{\cal L}
=\frac12(\partial_0\phi)^2+\frac12(\vec\partial\phi)^2.
\ee
The mode expansion for the field is 
\be \label{eq:ETmode}
\phi=\frac{1}{(2\pi)O^{3/2}}\int \frac{d^3p}{\sqrt{2|\vec p|}}
   \left\{ a(\vec{p})e^{-ip\cdot x} + a^\dagger(\vec{p})e^{ip\cdot x}\right\},
\ee
with the nonzero commutation relation
\be
[a(\vec{p}),a^\dagger(\vec{p}^{\,\prime})]=\delta(\vec{p}-\vec{p}^{\,\prime}).
\ee
Without normal ordering, the vacuum expectation value of the Hamiltonian
density is
\be
\langle0|{\cal H}|0\rangle_{\rm free}=\frac{1}{2(2\pi)^3}\int p\,d^3p,
\ee
which is, of course, infinite.  

If the field is constrained to
satisfy periodic boundary conditions at parallel plates,
separated a distance $a$ and perpendicular to the $z$ axis,
the $z$ component of momentum is constrained to be discrete
\be
\phi(z+a)=\phi(z)\;\;\Rightarrow\;\; e^{ip_z a}=1\;\;\Rightarrow\;\; p_z=\frac{2\pi n}{a},
\ee
with $n$ any integer.  The integral over $p_z$ is then replaced by a sum
\be  \label{eq:calE_ET}
\langle0|{\cal H}|0\rangle_{\rm PBC}
=\frac{2\pi}{a}\sum_n\frac{1}{2(2\pi)^3}\int E_n\,d^2p_\perp
   =\frac{1}{2a(2\pi)^2}\sum_{n=-\infty}^\infty\int E_n\,d^2p_\perp,
\ee
where $E_n\equiv\sqrt{p_\perp^2+\left(\frac{2\pi n}{a}\right)}$.
This density is also infinite, requiring regularization and subtraction.

To regulate, we introduce a heat-bath factor $e^{-\lambda E_n}$ and
take the limit of $\lambda\rightarrow0$ after computing the density
and subtracting the free-space density, computed with the same
regularization.  The density for periodic boundary conditions is then
\be
\langle0|{\cal H}|0\rangle_{\rm PBC}
   =\frac{1}{2a(2\pi)^2}\sum_{n=-\infty}^\infty\int E_n e^{-\lambda E_n}\,d^2p_\perp.
\ee
The integral is readily performed in polar coordinates where the angular
integral yields $2\pi$ and a change of variable from $p_\perp$ to $E_n$
leaves
\be
\int E_n e^{-\lambda E_n}\,d^2p_\perp=2\pi\int_{2\pi n/a}^\infty E_n^2 e^{-\lambda E_n} dE_n
=\frac{4\pi e^{-2 \lambda \pi n/a}}{\lambda^2a^2}(a^2+2 a \lambda n\pi+2\lambda^2n^2\pi^2).
\ee
The sum over $n$ is computed as a geometric series in $ e^{-2 \lambda \pi /a}$ or 
derivatives of such a series, to account for leading factors of $n$ and $n^2$.
The final result is
\be
\langle{\cal H}\rangle_{\rm PBC}=\frac{3}{2\pi^2\lambda^4}-\frac{\pi^2}{90 a^4}
                         +\frac{2\pi^4\lambda^2}{315 a^6}+{\cal O}(\lambda^4).
\ee
Repeating similar steps for the free-space density yields
\be
\langle0|{\cal H}|0\rangle_{\rm free}=\frac{3}{2\pi^2\lambda^4}.
\ee
Subtraction and the limit $\lambda\rightarrow0$ give the regulated vacuum energy 
density
\be
{\cal E}_{\rm PBC}\equiv\langle0|{\cal H}|0\rangle_{\rm PBC}-\langle0|{\cal H}|0\rangle_{\rm free}
                        =-\frac{\pi^2}{90 a^4}.
\ee
The energy per unit area between the plates is $a{\cal E}_{\rm PBC}$, and its derivative
yields the (negative of the) pressure
\be
P_{\rm PBC}=-\frac{d}{da}(a{\cal E}_{\rm PBC})=-\frac{\pi^2}{30a^4}.
\ee

If instead we impose boundary conditions were the field $\phi$ must be
zero at the plates, in analogy with the electromagnetic boundary conditions
at a perfect conductor, the $z$ component of momentum is discretized
as $p_z=n\pi/a$.  The calculation is then the same as for periodic
boundary conditions, with $a$ replaced by $2a$ except in the step of
finding the energy per unit area where $a$ is unchanged as the separation
between the plates.  We then have 
\be
{\cal E}_{\rm 0BC}=-\frac{\pi^2}{90 (2a)^4}
\ee
and
\be
P_{\rm 0BC}=-\frac{d}{da}(a{\cal E}_{\rm 0BC})=-\frac{\pi^2}{360a^4}.
\ee

These considerations are all at zero temperature.  To extend to finite
temperature $T$, we must use the free energy density for these bosons, obtained 
by~\cite{Plunien} replacing $\frac12E_n$ with $\frac12E_n+kT\ln[1-e^{-E_n/kT}]$
in Eq.~(\ref{eq:calE_ET}):
\be  \label{eq:calF_ET}
{\cal F}_{\rm PBC}=\frac{1}{2a(2\pi)^2}\sum_{n=-\infty}^\infty\int \{E_n+2kT\ln[1-e^{-E_n/kT}]\}\,d^2p_\perp.
\ee
From this we must subtract the free energy density without the periodic
boundary conditions
\be
{\cal F}_{\rm free}=\frac{1}{2(2\pi)^3}\int \{p+2kT\ln[1-e^{-p/kT}]\}\,d^3p.
\ee
The difference of the first terms is as before, after regularization.
The temperature dependent second terms do not require regularization; they
contribute
\be  \label{eq:calFreduced}
{\cal F}_{\rm PBC}^T=\frac{kT}{a(2\pi)^2}\sum_{n=-\infty}^\infty\int\ln[1-e^{-E_n/kT}]\,d^2p_\perp
                       -\frac{kT}{(2\pi)^3}\int\ln[1-e^{-p/kT}]\,d^3p
\ee
To simplify this expression, use new integration variables $x=E_n/kT$ and $y=p/kT$.  
Then, with $d^2p_\perp$ replaced by $2\pi (kT)^2 xdx$ and $d^3p$ replaced by
$4\pi(kT)^3 y^2 dy$, we have
\be
{\cal F}_{\rm PBC}^T=\frac{(kT)^3}{2\pi a}\sum_{n=-\infty}^\infty\int_{2\pi n/kTa}^\infty\ln[1-e^{-x}]\,x\,dx
                       -\frac{2(kT)^4}{(2\pi)^2}\int_0^\infty\ln[1-e^{-y}]\,y^2\,dy.
\ee
Given the integral representation for the zeta function
\be
\zeta(s)=\frac{1}{(s-1)!}\int_0^\infty\frac{x^{s-1}dx}{e^x-1}
\ee
and an integration by parts, this part of the free energy density can be rewritten as
\be
{\cal F}_{\rm PBC}^T=-\frac{(kT)^3}{2\pi a}\zeta(3)             
                     +2\frac{(kT)^3}{2\pi a}\sum_{n=1}^\infty\int_{2\pi n/kTa}^\infty\ln[1-e^{-x}]\,x\,dx
                       +\frac{4(kT)^4}{(2\pi)^2}\zeta(4).       
\ee

For low temperatures, where $2\pi a/kT\gg1$, the remaining logarithm can be expanded in
powers of $e^{-x}$.  The leading contribution is of order $e^{-2\pi n/kTa}$.  Therefore, the
dominant exponential contribution is for $n=1$; the subleading contribution
for $n=1$ is of order $e^{-4\pi/kTa}$, which is the same order as the leading
$n=2$ contribution.  The low temperature form of the Casimir free-energy density is then
\be
{\cal F}_{\rm PBC}=-\frac{\pi^2}{90 a^4}-\frac{(kT)^3}{2\pi a}\zeta(3) 
                     -2\frac{(kT)^3}{2\pi a}\left[1+\frac{2\pi}{kTa}\right]e^{-2\pi n/kTa}   
                       +\frac{4(kT)^4}{(2\pi)^2}\zeta(4).
\ee
For a high-temperature expansion, see \cite{Plunien}.

\subsection{\it Light-front quantization}

\begin{figure}[tb]
\begin{center}
\epsfig{file=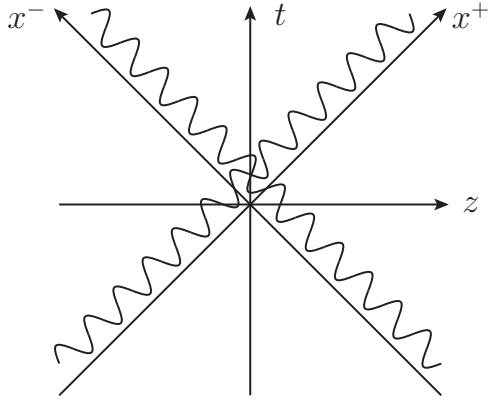,scale=0.75}
\caption{Light-front coordinates.  The light wave along the $x^-$ axis
is everywhere at one light-front time $x^+$.
\label{fig:litecone}}
\end{center}
\end{figure}

Light-front coordinates~\cite{Dirac,LFreview5} are the light-front time
$x^+\equiv t-z$, the longitudinal spatial coordinate
$x^-\equiv t+z$, and the transverse coordinates
$\vec{x}_\perp=(x,y)$.  The three light-front spatial
coordinates are combined as $\ub{x}=(x^-,\vec{x}_\perp)$.
The axes for light-front time and longitudinal space are 
shown in Fig.~\ref{fig:litecone}.  The conjugate light-front 
energy and momentum are $p^-\equiv E-p_z$ and
$\ub{p}=(p^+\equiv E+p_z,\vec{p}_\perp\equiv(p_x,p_y))$.
The scalar product of momentum and position four-vectors is
$p\cdot x=\frac12(p^=x^-+p^-x^+)-\vec{p}_\perp\cdot\vec{x}_\perp$.
The mass shell condition  $p^2=m^2$ can then be 
reinterpreted as $p^-=(p_\perp^2+m^2)/p^+$.  

\begin{figure}[ht]
\begin{center}
\epsfig{file=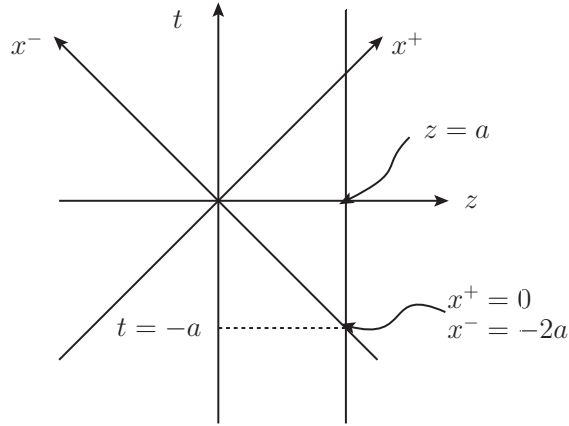,scale=0.75}
\caption{Parallel plates located at $z=0$ and $z=a$. 
The plate at $z=a$ intersects the $x^-$ axis at $x^-\equiv t-z=-2a$.
\label{fig:plates}}
\end{center}
\end{figure}

A calculation of the Casimir force in light-front coordinates~\cite{LFCasimir}
requires care with respect to the interpretation of energy
and the meaning of the boundary conditions.  The energy must be 
the ordinary equal-time energy $E$, not the light-front energy 
$P^-$, because the variation of $E$ is what yields the force. 
The plates must be at rest in some frame rather than separated 
by a fixed distance in $x^-$; in the latter case, the plates 
would be moving with the speed of light in any rest frame, which
is unphysical.   The correct configuration is depicted in Fig.~\ref{fig:plates}.
There is also the alternate possibility of plates separated in a 
transverse direction, which we consider in a subsequent section.

The Lagrangian for a massless scalar field is, again,
\be
{\cal L}=\frac12(\partial_\mu\phi)^2
=\frac12\partial_-\phi\partial_+\phi-\frac12(\senk{\partial}\phi)^2,
\ee
and the light-front Hamiltonian density is
\be
{\cal H}^-\equiv\partial_-\phi\frac{\delta{\cal L}}{\delta(\partial_-\phi)}-{\cal L}=\frac12(\senk{\partial}\phi)^2.
\ee
We will also need the light-front longitudinal momentum density
\be
{\cal H}^+=\partial_-\phi\frac{\delta{\cal L}}{\delta(\partial_+\phi)}
         =2\left(\frac{\partial\phi}{\partial x^-}\right)^2.
\ee
The mode expansion for the field is 
\be
\phi=\int_{p^+\geq0} \frac{d\ub{p}}{\sqrt{16\pi^3 p^+}}
   \left\{ a(\ub{p})e^{-ip\cdot x} + a^\dagger(\ub{p})e^{ip\cdot x}\right\},
\ee
with the nonzero commutation relation
\be
[a(\ub{p}),a^\dagger(\ub{p}')]=\delta(\ub{p}-\ub{p}')\equiv\delta(p^+-p^{\prime +})\delta(\senk{p}-\senk{p'}).
\ee

\subsubsection{\it Longitudinal separation \label{sec:longitudinal}}

For comparison with the equal-time calculation, we first consider the 
calculation for periodic boundary conditions consistent with plates perpendicular to 
the $z$ axis. The periodicity condition is $\phi(z+a)=\phi(z)$, but we must
translate this to light-front coordinates, where we have
\be
\phi(x^++a,x^--a,\vec{x}_\perp)=\phi(x^+,x^-,\vec{x}_\perp).
\ee
The discretization of the conjugate momentum $p^+$ is then given implicitly by
\be
-p^+a/2+p^-a/2=2\pi n,
\ee
with $p^-=p_\perp^2/p^+$ and $n$ any integer in the interval $(-\infty,\infty)$.
The solution of this constraint, that respects the positivity of $p^+$, is
\be \label{eq:p+discrete}
p_n^+\equiv \frac{2\pi}{a}n+\sqrt{\left(\frac{2\pi}{a}n\right)^2+p_\perp^2}.
\ee
The full range of $p^+$ is recovered, with $n=-\infty$ yielding $p^+=0$
and $n=\infty$, $p^+=\infty$.

With $p^+$ being discrete, the integration over $p^+$ is replaced by
a sum over $n$
\be
\int dp^+ =\int \frac{dp^+}{dn}dn\rightarrow \sum_n \frac{dp^+}{dn}=\frac{2\pi}{a}\sum_n\frac{p_n^+}{E_n}.
\ee
The mode expansion of the field is then
\bea  \label{eq:mode}
\phi(x^+=0)&=&\frac{1}{\sqrt{2a}}\sum_n \int \frac{d^2p_\perp}{2\pi\sqrt{E_n}}
   \left\{ a_n(\vec{p}_\perp)e^{-ip_n^+x^-/2+i\vec{p}_\perp\cdot\vec{x}_\perp} \right.  \\
  && \left. \rule{1.3in}{0mm}
     + a_n^\dagger(\vec{p}_\perp)e^{ip_n^+x^-/2-i\vec{p}_\perp\cdot\vec{x}_\perp}\right\},
     \nonumber
\eea
where we define discrete annihilation operators
\be
a_n(\vec{p}_\perp)=\sqrt{\left|\frac{dp^+}{dn}\right|}\;a(p_n^+,\vec p_\perp),
\ee
for which the commutation relation becomes
\be
[a_n(\vec{p}_\perp),a_{n'}^\dagger((\vec{p}_\perp)^{\,\prime})]
    =\delta_{nn'}\delta(\vec{p}_\perp-\vec{p}_\perp^{\,\prime}).
\ee
The leading $\frac{1}{\sqrt{2a}}$ factor in (\ref{eq:mode}) is the
normalization factor for the discrete basis functions
$e^{-ip_n^+x^-/2+i\vec{p}_\perp\cdot\vec{x}_\perp}$ on the interval
$-2a<x^-<0$.

The physical energy density is built from a sum of 
the vacuum expectation values of the light-front energy and
longitudinal momentum densities, which are
\bea
\langle0|{\cal H}^-|0\rangle&=&\frac{1}{4a}\sum_{n,n'}\int 
      \frac{d^2p_\perp d^2p'_\perp}{(2\pi)^2\sqrt{E_n E_{n'}}}
      \vec{p}_\perp\cdot\vec{p}_\perp^{\,\prime}
      \langle0|a_n(\vec{p}_\perp)a_{n'}^\dagger(\vec{p}_\perp^{\,\prime})|0\rangle \nonumber \\
      &=&\frac{1}{4a}\sum_n \int \frac{d^2p_\perp}{(2\pi)^2 E_n}p_\perp^2
\eea
and
\bea
\langle0|{\cal H}^+|0\rangle&=&\frac{2}{2a}\sum_{n,n'}\int 
      \frac{d^2p_\perp d^2p'_\perp}{(2\pi)^2\sqrt{E_n E_{n'}}} \frac{p_n^+ p_{n'}^+}{4}
      \langle0|a_n(\vec{p}_\perp)a_{n'}^\dagger(\vec{p}_\perp^{\,\prime})|0\rangle \nonumber \\
      &=&\frac{1}{4a}\sum_n \int \frac{d^2p_\perp}{(2\pi)^2 E_n} (p_n^+)^2.
\eea
The energy density is one-half of the sum of these
\bea
{\cal E}_{\rm LF}&\equiv&\frac12(\langle0|{\cal H}^-|0\rangle+\langle0|{\cal H}^+|0\rangle) \\
&=& \frac{1}{8a}\sum_n \int \frac{d^2p_\perp}{(2\pi)^2 E_n} (2E_n^2+2\frac{2\pi}{L}nE_n),
\eea
relative to light-front coordinates.  The second term is proportional
to $\sum_{n=-\infty}^\infty n=0$ and therefore zero itself.  We then obtain
\be
{\cal E}_{\rm PBC}^{\rm LF}=\frac{1}{4a}\sum_n\int \frac{d^2p_\perp}{(2\pi)^2}E_n.
\ee

However, this is not the same as the energy density relative to 
equal-time coordinates, which we denote simply by ${\cal E}_{\rm PBC}$.
We must first integrate over a finite separation between the plates
\be
\int_0^a dz {\cal E}_{\rm PBC}=\int_{-2a}^0 dx^- {\cal E}^{\rm LF}_{\rm PBC}.
\ee
After a change of variable from $x^-$ to $z=(x^+ + x^-)/2$ at fixed $x^+$ 
on the right hand side, this connection reduces to
\be
{\cal E}_{\rm PBC}=\frac{1}{a}\int_0^a 2dz {\cal E}^{\rm LF}_{\rm PBC}=2{\cal E}^{\rm LF}_{\rm PBC}.
\ee
Thus, the energy density is
\be
{\cal E}_{\rm PBC}=\frac{1}{2a}\sum_n\int \frac{d^2p_\perp}{(2\pi)^2}E_n,
\ee
which is identical with the equal-time result (\ref{eq:calE_ET}). 

It is regulated in the same way.  The heat-bath factor is again
$e^{-\lambda E_n}$ rather than an exponentiation of $\Pminus$, because
the heat bath should be a rest with the plates
rather than moving at the speed of light.  The subtraction is
also the same, because the energy density of the free-space vacuum
is independent of coordinates.

The case of zero boundary conditions is easily handled, once one
recognizes that the discretization of $p+$ in (\ref{eq:p+discrete})
corresponds directly to the quantization of $p_z$ as $2\pi n/a$.
The zero boundary conditions would then correspond to
$p_z=\pi n/a$ and 
\be 
p_n^+\equiv \frac{\pi}{a}n+\sqrt{\left(\frac{\pi}{a}n\right)^2+p_\perp^2}.
\ee
The analysis then proceeds in the same way as for periodic boundary
conditions, except that $E_n$ is given by $\sqrt{\left(\frac{\pi}{a}n\right)^2+p_\perp^2}$.
As for the equal-time calculation, the zero-boundary-condition 
result is obtained by replacing $a$ with $2a$ in the energy density.

\subsubsection{\it Transverse separation}  \label{sec:transverse}

For a transverse separation, again by a distance $a$, there is
less that is peculiar about a light-front formulation.  Without
loss of generality, we pick $x$ as the transverse direction.
The periodicity requirement is then
\be
\phi(x^+,x^-,x+a,y)=\phi(x^+,x^-,x,y).
\ee
This implies discretization in $p_x$ to values $p_n\equiv2\pi n/a$
The mode expansion becomes
\bea
\phi(x^+=0)&=&\frac{1}{\sqrt{a}}\sum_n \int \frac{dp^+ dp_y}{\sqrt{8\pi^2 p^+}}
   \left\{ a_n(p^+,p_y)e^{-ip^+x^-/2+ip_n x+ip_y y} \right.  \\
  && \left. \rule{1.3in}{0mm}
     + a_n^\dagger(p^+,p_y)e^{ip^+x^-/2-ip_n x -i p_y y}\right\},
     \nonumber
\eea
with discrete annihilation operators
\be
a_n(p^+,p_y)=\sqrt{\frac{2\pi}{a}}a(p^+,p_n,p_y),
\ee
that obey the commutation relation
\be
{[}a_n(p^+,p_y),a_{n'}^\dagger(p^{\prime +},p'_y]
=\delta_{nn'} \delta(p^+-p^{\prime +}) \delta(p_y-p'_y).
\ee
The leading factor $\frac{1}{\sqrt{a}}$ represents the normalization 
of the wave functions $e^{-ip^+x^-/2+ip_n x+ip_y y}$ on the interval $0<x<a$.

The light-front energy and longitudinal momentum densities are
\bea
\langle0|{\cal H}^-|0\rangle&=&\frac{1}{2a}\sum_{nn'}
  \int\frac{dp^+dp_ydp^{\prime+}dp'_y}{8\pi^2\sqrt{p^+p^{\prime+}}}
  (p_n p_{n'}+p_y p'_y) 
    \langle0|a_n(p^+,p_y)a_{n'}^\dagger(p^{\prime+},p'_y)|0\rangle \nonumber \\
  &=&\frac{1}{2a}\sum_n \int\frac{dp^+ dp_y}{8\pi^2} \frac{p_n^2+p_y^2}{p^+}
\eea
and
\bea
\langle0|{\cal H}^+|0\rangle&=&\frac{2}{a}\sum_{nn'}
   \int\frac{dp^+dp_ydp^{\prime+}dp'_y}{8\pi^2\sqrt{p^+p^{\prime+}}}
   \frac{p^+ p^{\prime +}}{4}\langle0|a_n(p^+,p_y)a_{n'}^\dagger(p^{\prime+},p'_y)|0\rangle \nonumber \\
     &=&\frac{1}{2a}\sum_n \int\frac{dp^+ dp_y}{8\pi^2} p^+.
\eea
When summed, they yield
\be
{\cal E}^{\rm LF}_{\rm PBC}=\frac{1}{2a}\sum_n \int \frac{dp^- dp^+ dp_y}{8\pi^2}
   \frac{p^- + p^+}{2} \delta\left(p^--\frac{p_n^2+p_y^2}{p^+}\right),
\ee
where the delta function enforces the mass-shell condition and can be
rewritten as
\be
\delta\left(p^--\frac{p_n^2+p_y^2}{p^+}\right)=p^+\delta(p^2)=p^+\delta(E^2-E_n^2),
\ee
with $E_n=\sqrt{\left(\frac{2\pi n}{a}\right)^2 +p_z^2+p_y^2}$.
The integral is then trivially converted to an integral with respect to
equal-time variables $E=(p^+ +p^-)/2$ and $p_z=(p^+-p^-)/2$:
\be
{\cal E}^{\rm LF}_{\rm PBC}=\frac{1}{2a} \sum_n \int \frac{2dE dp_z dp_y}{8\pi^2}
   E(E+p_z)\frac{1}{2E_n}\delta(E-E_n).
\ee
The $p_z$ term integrates to zero, being odd in $p_z$, and it is this
term that would be missed if only the minus density $\langle0|{\cal H}^-|0\rangle$
was used to represent the energy; the plus contribution, which was critical
in the longitudinal case, is zero in the transverse case.

We have thus found the energy density relative to light-front coordinates to
be given by
\be
{\cal E}^{\rm LF}_{\rm PBC}=\frac{1}{4a}\sum_n\int\frac{dp_z dp_y}{(2\pi)^2}E_n.
\ee
Just as for the longitudinal case, the energy density relative to equal-time
coordinates is obtained with multiplication by two, to find
\be
{\cal E}_{\rm PBC}=\frac{1}{2a}\sum_{n=-\infty}^\infty\int\frac{dp_z dp_y}{(2\pi)^2} E_n,
\ee
which matches the usual equal-time result (\ref{eq:calE_ET}) and is of the same form
as in the longitudinal case. The use of zero boundary conditions again alter
the result only by changing the discretization to $p_n=\pi n/a$.

The free energy density in light-front coordinates is identical to the
equal-time form (\ref{eq:calF_ET}), and no additional calculation is necessary to obtain
the finite-temperature contributions.  This happens 
because the discrete spectrum $E_n$ and the Boltzmann factor $e^{-E_n/kT}$ are
the same.  The direct equality of the spectra is explicit.  The choice of 
Boltzmann factor is driven by the physics.  A heat bath at temperature $T$
should be at rest~\cite{Elser,SDLCQ,Strauss:2008zx,Strauss:2009uj}; a 
light-front Boltzmann factor of the form $e^{-P^-/kT}$ would correspond 
to a heat bath moving with the speed of light.

\subsubsection{\it Light-like boundary conditions \label{sec:lightlike}}

The first attempt to compute the Casimir force in light-front coordinates
was by Lenz and Steinbacher~\cite{Lenz}.  In the longitudinal direction
they applied light-like periodic boundary conditions on
the scalar field $\phi$,
\be 
\phi(x^+,x^-+a,x,y)=\phi(x^+,x^-,x,y),
\ee
and studied the vacuum expectation value of the light-front energy.
For this expectation value they obtained
\be
\langle\Pminus\rangle=\frac{1}{2a}\int\frac{d^2 k_\perp}{(2\pi)^2}
   \sum_{n=0}^\infty\omega_n(k_\perp) e^{-\lambda^-2\pi n/a-\lambda^+\omega_n(k_\perp)},
\ee
with light-front energies $\omega_n(k_\perp)\equiv \frac{k_\perp^2}{4\pi n/a}$. The $\lambda^\pm$
regulate the $k^+=2\pi n/a$ and $\omega_n$ dependencies separately.  Their
computation of the sum yields
\be \label{eq:Lenz}
\langle\Pminus\rangle=\frac{1}{8\pi^2(\lambda^+\lambda^-)^2}-\frac{1}{24\lambda^{+2}a^2}
   +\frac{\pi^2}{120 a^4}\left(\frac{\lambda^-}{\lambda^+}\right)^2.
\ee
The regulator and separation dependence do not separate, making interpretation
difficult.

For a transverse separation, their calculation for periodic boundary 
conditions leads to the standard result.  As noted above, this success
is due to the lack of a contribution from the vacuum expectation value
of the longitudinal light-front momentum to the ordinary energy, allowing
use of only $\langle\Pminus\rangle$ to give the full answer.

They also studied what happens when the longitudinal boundary condition
is replaced with a boundary condition near the light front
\be 
\phi(x^+,x^-+a,x+sa,y)=\phi(x^+,x^-,x,y),
\ee
with $s$ taken to approach zero. This condition can, of course, be
transformed into an equivalent transverse boundary condition for
any nonzero $s$.  For such a condition, the light-front calculation
yields the correct result.  This approach leads naturally to 
consideration of modified light-front coordinates known as
oblique light-front coordinates~\cite{Weldon}.

\subsection{\it Oblique light-front coordinates \label{sec:oblique}}

The application of oblique light-front coordinates to the Casimir-force problem
was considered by Almeida, {\em et al.}~\cite{Almeida}.  The coordinates are
defined as
\be
\bar{x}^0=t+z,\;\;\bar{x}^3=z,\;\;\bar{x}=x,\;\;\bar{y}=y,
\ee
with the chosen time coordinate $\bar{x}^0$ equivalent to the light-front time $x^+$.
For a massless scalar field $\phi$, the Lagrangian density is
\be
{\cal L}_{\rm oblique}=-\bar{\partial}_0\phi\bar{\partial}_3\phi-\frac12(\bar{\partial}_\perp\phi)^2
                       -\frac12(\bar{\partial}_3\phi)^2.
\ee
The mode expansion for the field is
\be
\phi=\frac{1}{(2\pi)^{3/2}}\int d^2\bar{k}_\perp\int_0^\infty \frac{d\bar{k}_3}{2\bar{k}_3}
         \left[e^{-i\tilde{k}\cdot\bar{x}}a(\bar{k})
                +e^{i\tilde{k}\cdot\bar{x}}a^\dagger(\bar{k})\right],
\ee
with $\bar{k}$ the conjugate four-momentum, 
$\tilde{k}\equiv(\bar{k}_0, -\bar{k}_3,-\bar{k}_\perp)$,
and
\be
\bar{k}_0=\frac{\bar{k}_3^2+\bar{k}_\perp^2}{2\bar{k}_3}.
\ee
The nonzero commutation relation for the creation and annihilation operators is
\be
[a(\bar{k}),a^\dagger(\bar{q})]=2\bar{k}_3\delta(\bar{k}-\bar{q}).
\ee
The Hamiltonian density is then
\be
{\cal H}=\frac12\int d^2\bar{k}_\perp\int \frac{d\bar{k}_3}{2\bar{k}_3}
        \bar{k}_0[a(\bar{k})a^\dagger(\bar{k})+a^\dagger(\bar{k})a(\bar{k})].
\ee

As for the other methods, a transverse boundary condition yields the 
correct result and only the longitudinal boundary condition requires care.
If the longitudinal condition is a simple periodicity in $\bar{x}^3=z$
\be 
\phi(\bar{x}^0,\bar{x}^3+a,\bar{x}_\perp)=\phi(\bar{x}^0,\bar{x}^3,\bar{x}_\perp),
\ee
the plates are at rest in an inertial frame, unlike an $x^-$ separation,
because the condition is applied to the $z$ direction.  However, the 
correct result is not obtained; the regulator dependence and separation
dependence remain entangled, just as in the light-front expression (\ref{eq:Lenz}).
The difficulty with this longitudinal condition is that the periodicity is
taken at different Minkowski times~\cite{Almeida}. 

To avoid this inconsistency, a different longitudinal condition is used
\be 
\phi(\bar{x}^0+a,\bar{x}^3+a,\bar{x}_\perp)=\phi(\bar{x}^0,\bar{x}^3,\bar{x}_\perp),
\ee
one that is fully equivalent to the longitudinal case in equal-time 
coordinates.\footnote{This is the direct inspiration for the work
of Chabysheva and Hiller~\protect\cite{LFCasimir}, that use of
oblique coordinates is not necessary but rather that one must
make the correct choice of longitudinal boundary condition.}  This condition
leads to the discretization
\be
\frac{\bar{k}_\perp^2-\bar{k}_3^2}{2\bar{k}_3}=\frac{2\pi n}{a},
\ee
with $n=0,\pm1,\pm2,\ldots$.  The regulated vacuum expectation value of the
Hamiltonian density is
\be
\langle{\cal H}\rangle=\frac{1}{2a}\frac{1}{(2\pi)^2}\sum_{n=-\infty}^\infty
  \int d^2\bar{k}_\perp\int_0^\infty d\bar{k}_3\,\bar{k}_0
  \delta\left(\frac{\bar{k}_\perp^2-\bar{k}_3^2}{2\bar{k}_3}-\frac{2\pi n}{a}\right)
  e^{-\lambda^3\bar{k}_3-\lambda^0\bar{k}_0}.
\ee
A change of variables to $k_3\equiv\bar{k}_3-E_k$ and $\vec{k}_\perp=\vec{\bar{k}}_\perp$,
with $E_k\equiv\sqrt{k_\perp^2+k_3^2}$, and use of the $\delta$ function to perform the
$k_3$ integral, leaves
\be
\langle{\cal H}\rangle=\frac{1}{2a}\frac{1}{(2\pi)^2}\sum_{n=-\infty}^\infty
  \int d^2k_\perp \, E_k e^{\lambda^3 2\pi n/a-(\lambda^0+\lambda^3)E_k}.
\ee
Evaluation of the sum and integral then give~\cite{Almeida}
\be
\langle{\cal H}\rangle=\frac{1}{4\pi(\lambda^3\lambda^0)^2}
    -\frac{1}{32\pi^2(\lambda_3)^4}
    +\frac{3\lambda^0}{64\pi^2(\lambda^3)^5}
    -\frac{3(\lambda^0)^2}{64\pi^2(\lambda^3)^6}
    -\frac{\pi^2}{90a^4}+{\cal O}(\frac{\lambda^2}{a^6},\frac{1}{\lambda^4}).
\ee
This may differ in the regulator dependence, but this dependence now separates
from the $a$ dependence,
leaving the correct $-\pi^2/90a^4$ term as the physical
contribution to the Casimir energy.

The correction for nonzero temperature $T$ is computed from
the free energy density~\cite{Rodrigues}
\be
{\cal F}_{\rm PBC}^T=\frac{kT}{a(2\pi)^2}\sum_{n=-\infty}^\infty\int d^2\bar{k}_\perp d\bar{k}_3
      \ln\left[1-e^{-\bar{k}_0/kT}\right]\delta\left(\frac{\bar{k}_\perp^2-\bar{k}_3^2}{2\bar{k}_3}-\frac{2\pi n}{a}\right),
\ee
where, as before, the $\delta$ function enforces the discretization imposed by the periodic
boundary conditions.  Use of the same change of variables and integration over the
delta function reduces this expression to
\be
{\cal F}_{\rm PBC}^T=\frac{kT}{a(2\pi)^2}\sum_{n=-\infty}^\infty\int d^2\bar{k}_\perp\ln\left[1-e^{-E_k/kT}\right],
\ee
with $E_k$ discretized as $E_k=\sqrt{k_\perp^2+\left(\frac{2\pi n}{a}\right)^2}$.  This is the same as the 
(unsubtracted) equal-time expression (\ref{eq:calFreduced}).

\section{Summary} \label{sec:summary}

From cosmological to atomic scales, the quantum vacuum plays a 
significant role in our understanding of the Universe. As suggested
by Davies~\cite{Davies}, ``the vacuum holds the key to 
a full understanding of the forces of nature.''  The Casimir force
is a prime example; though it can be computed from interatomic
forces, it can also be represented in terms of variations in
the quantum vacuum energy density.

Any system that can be analyzed with more than one coordinate
system will be better understood.  Here we have shown how the
Casimir force, including thermal effects, can be computed in 
light-front quantization and that the results agree with
those from equal-time quantization, including modest extensions
of previous work~\cite{LFCasimir} from periodic to zero boundary
conditions and to finite temperature.  This can be an aid to
the incorporation of vacuum effects into nonperturbative
light-front calculations~\cite{LFreview5}, where the vacuum
has traditionally been considered trivial.

The formulation of the Casimir boundary conditions in terms
of light-front coordinates will have applications in the
study of effective potentials between static sources.  One
immediately recognizes that a static source, meaning a 
source at rest in an inertial frame, will be moving in the
light-front coordinate $x^-$.  A light-front static source
model must then accommodate a moving source.  Taking the
notion of static source too literally, that is fixed in $x^-$,
would mean a source moving with the speed of light.  This
is the analog of the care taken here, that the parallel plates
of the Casimir effect are static in an inertial frame.

\section*{Acknowledgments}
This manuscript is dedicated to the memory of 
Joseph Sucher (1930-2019), thesis advisor for JRH.
The work by SSC and JRH was supported in part by 
the US Department of Energy through Contract No.\ DE-FG02-98ER41087
and in part by the Minnesota Supercomputing Institute
of the University of Minnesota with grants of computing
resources.


\end{document}